\documentclass[10pt,a4paper,twocolumn]{article}

\usepackage[utf8]{inputenc}
\usepackage[T1]{fontenc}
\usepackage{lmodern}          % FIX-F1: microtype font expansion fails without scalable T1 fonts
\usepackage{amsmath,amssymb,amsthm}
\usepackage{stmaryrd}
\usepackage{graphicx}
\usepackage{booktabs}
\usepackage{microtype}
\usepackage[margin=1.8cm]{geometry}
\usepackage{xurl}
\usepackage[hidelinks]{hyperref}
\usepackage{cite}
\usepackage{caption}

\let\oldunderscore\_
\renewcommand{\_}{\oldunderscore\allowbreak}

\newtheorem{definition}{Definition}

\title{\textbf{Benchmarking Flat Non-Clifford Hypergraph States, Antipodal Obstructions, and Contextuality Games on IBM Heron~r2 and Trapped-Ion Processors}}

\author{
  Filipe Martins Vono\thanks{Centro Universit\'ario de Belo Horizonte (UniBH), Belo Horizonte, MG, Brazil. ORCID: \href{https://orcid.org/0009-0009-4994-8533}{0009-0009-4994-8533}. Email: \texttt{contatofilipevono@gmail.com}},
  \quad Val\'eria Gonzaga Santos\thanks{Independent researcher, Belo Horizonte, MG, Brazil. Email: \texttt{gonzagasantosva@gmail.com}},
  \quad Andr\'e Oliveira Naufel de Toledo\thanks{Centro Universit\'ario de Belo Horizonte (UniBH) / \^Anima Educa\c{c}\~ao, Belo Horizonte, MG, Brazil. Email: \texttt{andre.toledo@animaeducacao.com.br}}
}

\date{December 2026}

\begin{document}
\maketitle

\begin{abstract}
We report 20 auditable hardware jobs across three processors---17 jobs (3.21~Resource Units) on two 156-qubit IBM Heron~r2 superconducting chips (\texttt{ibm\_fez}, \texttt{ibm\_kingston}) and 3 jobs on a 36-qubit all-to-all trapped-ion processor (\texttt{IonQ Forte Enterprise})---benchmarking compilation and circuit fidelity for non-Clifford hypergraph states $\lvert\psi_f\rangle$ with bent phase polynomials $f$, alongside Clifford and non-Clifford contextuality tests. Our sharpest controlled result is a same-job compilation comparison on \texttt{ibm\_fez} for an overlapping $3\times\mathrm{CCCZ}$ 16-qubit mirror circuit ($H^{\otimes n}U_f^\dagger U_f H^{\otimes n}$): replacing the default 16-qubit compilation (222 CZ, depth 515) with a calibration-selected 23-qubit zero-SWAP ancilla+RCCX synthesis (110 CZ, depth 119) raises raw mirror return by $+16.67$~pp, from $4/4{,}096$ ($0.10\%$) to $687/4{,}096$ ($16.77\%$), and to $592/2{,}576$ ($22.98\%$; unconditional joint yield $592/4{,}096=14.45\%$) post-selected on ancilla return. For $X$-basis flatness tests ($n=6$--$10$), bent states match uniform noise while non-flat controls reach nonuniformity ratio $R=\mathrm{TVD}_{\mathrm{hw}}/\mathrm{TVD}_{\mathrm{ideal}}=0.57$--$0.90$ (a descriptive nonuniformity metric insensitive to support permutations). On the Mermin--Peres magic square we obtain raw circuit win rates of $96.85\pm0.09\%$ on \texttt{ibm\_fez} (fitted readout correction $+1.50$~pp) and $97.29\pm0.45\%$ cluster-robust per-shot SE ($1{,}401/1{,}440$ plays, Row~1 $99.17\%$; difference $+0.45$~pp, 95\% shot-model CI $[-0.45,+1.34]$~pp) on \texttt{IonQ Forte Enterprise} using three parallel 12-qubit jobs with a fixed Alice/Bob logical partition and zero cross-partition SWAPs---noting explicitly that assigning each column to a fixed block yields a $100\%$ ($9/9$) classical ceiling under block-dependent strategies. At $n=30$, an XY4 GHZ phase scan gives a fitted parity harmonic $V=0.1560\pm0.0110$ ($\phi_0=-70.9^\circ$), and exact restricted local ceilings on our 8-input GHZ runs are $7/8$ ($n=6,10$) and $8/8$ ($n=16,20,30$). On the three-qubit CCZ state, Mermin-type ($83.95\%$, ideal $87.99\%$) and Hardy-type ($77.15\%$) games exceed the $75\%$ theoretical reference on a single device; on a connected 16-qubit degree-4 circuit, 16 target-generator tests average $84.39\%$ ($83.64\%$ on the three bridge-sensitive generators), but we construct an explicit half-separable state with local $Z$ flips that reproduces all 16 archived marginals without entanglement across the bridge. Finally, two-copy Bell sampling gives $\mathcal{W}_2=\mathcal{M}_{2}-2S_2$ of $-0.074\pm0.019$ and $-0.009\pm0.024\,$bits for Clifford controls versus $+1.386\pm0.028$ ($n=4$ CCZ), $+1.237\pm0.043$ ($n=4$ CCCZ), and $+1.245\pm0.055\,$bits ($n=6$ flat CCZ; descriptive bootstrap SEs); under factorized stabilizer mixtures $\rho\otimes\sigma$ and trusted Bell measurements, analytic H\"older ceilings are $\mathcal{W}_2 < 0.6655\,$bits ($n=4$) and $<1.1960\,$bits ($n=6$), so the $n=4$ points exceed the bound under that model while $n=6$ is inconclusive.
\end{abstract}

% =========================================================================
\section{Introduction}
% =========================================================================

Contextuality and nonlocal games give operational, tomography-free tests of nonclassical correlations~\cite{mermin1990simple,mermin1990extreme,peres1990incompatible,ghsz1990,brassard2005recasting,bravyi2018quantum}. Google Quantum AI recently reported a magic-square win probability of $0.9830\pm0.0001$, a Kochen--Specker--Bell violation, and GHZ parity games exceeding the classical value up to $N=71$ qubits on superconducting hardware~\cite{google2026contextuality}. % FIX-I1: Google GHZ result must be acknowledged
These experiments use stabilizer states, which are efficiently classically simulable.

Hypergraph states~\cite{rossi2013quantum,qu2013encoding,guhne2014entanglement} replace CZ edges by multi-controlled-$Z$ hyperedges, corresponding to phase polynomials $f:\mathbb{F}_2^n\to\mathbb{F}_2$ of algebraic degree $d\ge 3$. They violate local realism strongly~\cite{guhne2014entanglement,gachechiladze2016extreme}, are resources for Pauli-universal measurement-based computation~\cite{miller2016hierarchy,takeuchi2019universality}, and four-qubit instances have been prepared and shown to violate Mermin inequalities on a photonic chip~\cite{huang2024hypergraph}. % FIX-L1: prior experiments
When $f$ is bent~\cite{rothaus1976bent,carlet2021boolean}, the $X$-basis distribution of $\lvert\psi_f\rangle$ is exactly uniform; this connection underlies hidden-shift algorithms for bent functions~\cite{rotteler2010hidden}, which were used as early cross-platform hardware benchmarks~\cite{linke2017comparison}. % FIX-L2

In companion theoretical work~\cite{vono2026modular,vono2026antipodal} we studied (i) modular certification of bentness and (ii) the role of the ``antipodal'' CZ orbit in Maiorana--McFarland and rotation-symmetric bent constructions.
The present paper is an \emph{experimental benchmark}; it does not test, and cannot test, modular arithmetic statements, which are purely classical. Its contributions are:
\begin{itemize}
    \item \textbf{Flatness controls ($n=6$--$10$).} We show that flat (bent) states are uninformative under $X$-basis sampling, compute exact 4,096-shot uniform TVD references, and report the non-flat nonuniformity ratio $R=\mathrm{TVD}_{\mathrm{hw}}/\mathrm{TVD}_{\mathrm{ideal}}$.
    \item \textbf{Mirror circuits and zero-SWAP synthesis ($n=6$--$16$).} Return probabilities of $U_f^\dagger U_f$ up to 230 CZ gates on two processors, plus a same-job A/B comparison showing a $+16.67$~pp raw mirror-return gain ($4/4{,}096$ to $687/4{,}096$; $22.98\%$ post-selected) via a calibration-selected 23-qubit zero-SWAP ancilla+RCCX compilation.
    \item \textbf{Clifford games (cross-platform).} Magic square on IBM Heron~r2 (raw $96.85\pm0.09\%$) and all-to-all \texttt{IonQ Forte Enterprise} (raw $97.29\pm0.45\%$ cluster-robust SE, fixed Alice/Bob logical partition, zero cross-partition SWAPs; with an explicit disclosure that block position encodes the column and yields a $100\%$ block-specific classical ceiling), plus GHZ parity circuits ($n=6$--$30$) with exact restricted classical ceilings for the sampled inputs, run-to-run variability, a dynamical-decoupling ablation, and a 30-qubit fitted parity harmonic scan.
    \item \textbf{Non-Clifford tests.} Mermin- and Hardy-type games on the CCZ state, 16 target-generator tests on a connected 16-qubit degree-4 circuit (together with an explicit half-separable counterexample reproducing all 16 archived marginals), and a Bell-derived $\mathcal{W}_2$ diagnostic audited against exact $n$-qubit H\"older stabilizer-mixture ceilings under factorized copies.
\end{itemize}
All 20 hardware jobs---17 on IBM Quantum (3.21~Resource Units account-balance difference, $10.00 - 6.79$~RU) and 3 on \texttt{IonQ Forte Enterprise} via Open Quantum---are documented with job IDs, IBM summary JSONs, raw 12-qubit IonQ histograms and OpenQASM circuits, experiment/analysis scripts, and a pure-Python offline verification and figure-replay script~\cite{vono2026archive} (Table~\ref{tab:jobs}).

% =========================================================================
\section{Theoretical Framework}
% =========================================================================

\subsection{Hypergraph states, bentness and antipodal obstructions}

For $f:\mathbb{F}_2^n\to\mathbb{F}_2$, $\lvert\psi_f\rangle=U_fH^{\otimes n}\lvert0^n\rangle=2^{-n/2}\sum_x(-1)^{f(x)}\lvert x\rangle$ with $U_f=\sum_x(-1)^{f(x)}\lvert x\rangle\langle x\rvert$. Measuring in the $X$ basis yields $a\in\mathbb{F}_2^n$ with $p_X(a)=W_f(a)^2/2^{2n}$, where $W_f(a)=\sum_x(-1)^{f(x)\oplus a\cdot x}$.

\begin{definition}[Flat hypergraph state]
For even $n$, $\lvert\psi_f\rangle$ is flat iff $f$ is bent, $\lvert W_f(a)\rvert=2^{n/2}$ for all $a$, i.e.\ $p_X\equiv2^{-n}$.
\end{definition}

\noindent\textbf{Remark (what $X$-basis flatness can and cannot show).}
The uniform distribution is also produced by the maximally mixed state and by $\lvert0^n\rangle$. A small measured TVD of a bent state therefore does \emph{not} certify coherence; noise drives every distribution towards uniform. Moreover, with $S=4{,}096$ shots an \emph{exactly} uniform source has exact expected empirical TVD $(1-2^{-n})\Pr[\mathrm{Bin}(S-1,2^{-n})=\lfloor S/2^n\rfloor]$, equal to $0.049$, $0.099$, and $0.195$ at $n=6,8,10$ (and $0.939$ at $n=16$). Only non-flat states, whose ideal TVD is large, carry informative nonuniformity contrast, though TVD from uniform is invariant under arbitrary permutations of output labels: in our obstructed MM controls (whose ideal support occupies $1/4$ of all $2^n$ outcomes), a label permutation can yield $R=1$ with zero probability mass on the true Walsh support. Thus $R$ is strictly a descriptive nonuniformity ratio rather than a target-structure estimator; a sharper metric on the same circuits is the fraction of shots landing inside the predicted Walsh support.

For odd $n$ the maximum nonlinearity at $n=7$ is $56$~\cite{mykkeltveit1980covering}. % FIX-M2: "bent-concatenation bound" wording was wrong

We use two bent families:
\begin{itemize}
    \item \textbf{Maiorana--McFarland (MM)}~\cite{mcfarland1973family,dillon1974thesis}: $n=2m$, $f(y,x)=x\cdot Ay\oplus g(y)$ with $y=(0,\dots,m-1)$, $x=(m,\dots,2m-1)$, $A\in GL_m(\mathbb{F}_2)$ (here $A=I$), and $\deg g\in\{3,4\}$ acting on the $y$ register. The bilinear part is the \emph{antipodal CZ orbit} $\{(i,i+m)\}$.
    In our obstructed controls ($n=6,8,10$), removing the antipodal edge $(0,m)$ reduces $\mathrm{rank}(A)$ to $m-1$ (decoupling $x_0$), while adding the nearest-neighbour edge $(0,1)$ merely shifts $g(y)\mapsto g(y)\oplus y_0y_1$ inside the $y$ register; hence $p_X$ is supported on $2^{n-2}$ outcomes with $p_X=2^{2-n}$ and $\mathrm{TVD}_{\mathrm{ideal}}=3/4$ for every $n$. That rank-deficient MM functions are not bent is classical~\cite{dillon1974thesis,carlet2021boolean}.
    \item \textbf{Rotation-symmetric bent (RSB), $n=6$, degree 3}~\cite{stanica2008rotation,gao2012rotation}: the orbit sum of $x_0x_3$, $x_0x_1x_2$ and $x_0x_2x_4$ (11 monomials). Its control removes the \emph{entire} antipodal orbit (to preserve rotation symmetry), giving $\mathrm{TVD}_{\mathrm{ideal}}=7/8$. % FIX-M5
\end{itemize}

\subsection{Modular Walsh certification and the mod-3 fooler}

At $n=8$, $f_{\mathrm{fool}}=x_0x_1x_2x_3\oplus x_4x_5\oplus x_6x_7$ has $W\in\{\pm8,\pm56\}$, hence $W(a)^2\equiv 2^8\equiv1\pmod 3$ for all $a$, although $f_{\mathrm{fool}}$ is not bent ($\mathrm{TVD}_{\mathrm{ideal}}=45/64\approx0.703$). % FIX-M6: exact value; "universal" removed
We note that $W^2\equiv 2^n \pmod 3$ holds for \emph{any} $W\not\equiv0\pmod 3$, so a mod-3 test is weak for elementary reasons; the stronger mod-17 statements are proved in~\cite{vono2026modular}. On hardware, $f_{\mathrm{fool}}$ is simply a non-flat control with a different spectrum.

\subsection{Classical values of the games}\label{sec:games}

\begin{enumerate}
    \item \textbf{Magic square (4 qubits).} Classical value $8/9$; quantum value $1$~\cite{mermin1990simple,peres1990incompatible,cabello2001bell}.
    \item \textbf{GHZ parity game ($n$ qubits).} Inputs $x\in\{0,1\}^n$ of even weight; win iff the output parity equals $\mathrm{wt}(x)/2 \bmod 2$. For the \emph{uniform} input distribution the classical value is $\tfrac12+2^{-\lceil n/2\rceil}$~\cite{brassard2005recasting}. However, when only a fixed subset of 8 inputs is sampled repeatedly, a deterministic local strategy can exploit that fixed subset: exhaustive enumeration over all affine local strategies $b\oplus a\cdot x$ gives exact restricted local ceilings of $7/8$ ($87.5\%$) on our 8 sampled inputs at $n=6,10$, $8/8$ ($100\%$) at $n=16,20,30$, and $2/2$ ($100\%$) for the 2-input $n=50$ test. We therefore report both the restricted ceiling and the uniform-input theoretical reference, without fixed-question significance tests against the uniform value.
    \item \textbf{Three-qubit CCZ games.} On $\mathrm{CCZ}\lvert{+}{+}{+}\rangle$ each qubit chooses one of two dichotomic measurements. Enumerating the $2^6$ deterministic local strategies gives a theoretical local-strategy reference of $3/4$ for both (a)~a Mermin-type game with fixed target parities $(0,0,0,1)$ across contexts $(011,101,110,000)$, ideal $M=+3.0389$ and win rate $(4+M)/8=87.99\%$; and (b)~a Hardy-type game on contexts $XZZ$, $ZXZ$, $ZZX$ (win iff $x_i=z_jz_k$) and $XXX$ (win iff $x_0+x_1+x_2=2$), ideal $51/64=79.69\%$~\cite{hardy1993nonlocality,guhne2014entanglement,gachechiladze2016extreme}.
    \item \textbf{Generalized-stabilizer checks ($n=6$--$16$).} Measuring qubit $j$ in $X$ and its hypergraph neighbourhood in $Z$ tests one target generator $K_j=U_fX_jU_f^\dagger$. These are \emph{single-context} checks, not games: a classical distribution can pass any single check with probability one. In the earlier $n=6$--$16$ checks each test involved 4--5 qubits; in the full 16-generator test below, the three cross-wing bridge generators $K_3, K_4, K_5$ involve 7, 6, and 6 qubits respectively.
\end{enumerate}

% =========================================================================
\section{Hardware Protocol}
% =========================================================================

All experiments used Qiskit~2.6.2 and \texttt{qiskit-ibm-runtime}~0.52.1 (\texttt{SamplerV2}) on \texttt{ibm\_fez} and \texttt{ibm\_kingston} (native gates CZ, $R_Z$, $\sqrt{X}$, $X$), with 4,096 shots/circuit unless stated. Transpilation used \texttt{optimization\_level=3} (fixed seeds on \texttt{ibm\_fez} scripts; default seeds on early \texttt{ibm\_kingston} runners). Dynamical decoupling (DD) was off unless stated.

Table~\ref{tab:topology} lists transpiled CZ counts for four coupling maps (\texttt{opt=3}, single fixed seed). These are compiler outputs, not hardware results, and CZ counts are not directly comparable across architectures with different native entanglers and fidelities.

\begin{table}[t]
\centering
\caption{Transpiled CZ count (\texttt{opt=3}, single seed) relative to all-to-all connectivity.}
\label{tab:topology}
\resizebox{\columnwidth}{!}{%
\begin{tabular}{@{}lcccc@{}}
\toprule
State & All-to-All & 2D Grid & Heavy-Hex & 1D Ring \\
\midrule
$n=6$, deg 3  & 9  (1.00$\times$) & 12 (1.33$\times$) & 12 (1.33$\times$) & 13 (1.44$\times$) \\
$n=8$, deg 3  & 28 (1.00$\times$) & 38 (1.36$\times$) & 44 (1.57$\times$) & 47 (1.68$\times$) \\
$n=8$, deg 4  & 24 (1.00$\times$) & 35 (1.46$\times$) & 35 (1.46$\times$) & 41 (1.71$\times$) \\
$n=10$, deg 3 & 35 (1.00$\times$) & 55 (1.57$\times$) & 62 (1.77$\times$) & 74 (2.11$\times$) \\
$n=12$, deg 3 & 42 (1.00$\times$) & 62 (1.48$\times$) & 76 (1.81$\times$) & 91 (2.17$\times$) \\
$n=12$, deg 4 & 96 (1.00$\times$) & 150 (1.56$\times$) & 164 (1.71$\times$) & 200 (2.08$\times$) \\
\bottomrule
\end{tabular}%
}
\end{table}

\begin{figure*}[t]
    \centering
    \includegraphics[width=0.96\textwidth]{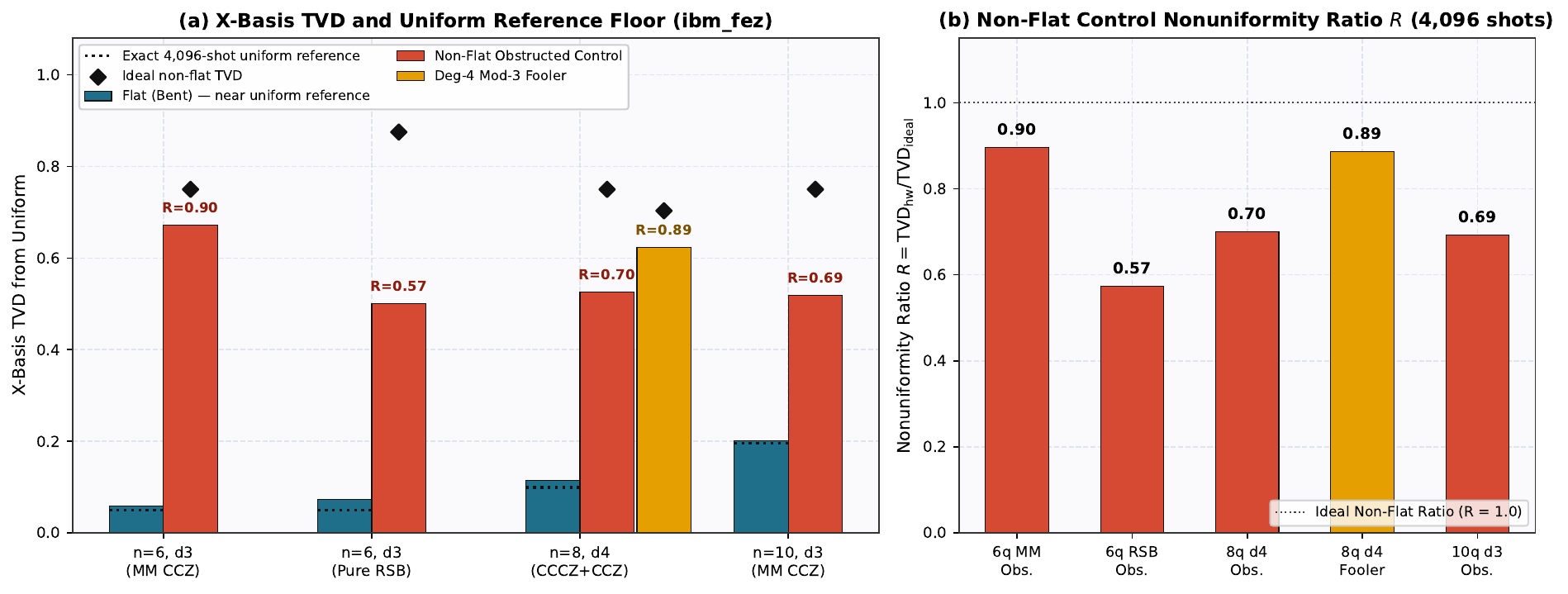}
    \caption{(a)~$X$-basis TVD from uniform on \texttt{ibm\_fez} (4,096 shots), showing exact 4,096-shot uniform references ($0.049, 0.099, 0.195$; dotted lines) and ideal non-flat TVDs (diamonds). Flat states lie near the uniform reference (with $n=6$ RSB and $n=8$ Fez exceeding uniform shot noise); panel~(b) summarizes the non-flat nonuniformity ratio $R=\mathrm{TVD}_{\mathrm{hw}}/\mathrm{TVD}_{\mathrm{ideal}}$ (Table~\ref{tab:flatness}).}
    \label{fig:fig1}
\end{figure*}

\begin{figure*}[t]
    \centering
    \includegraphics[width=0.96\textwidth]{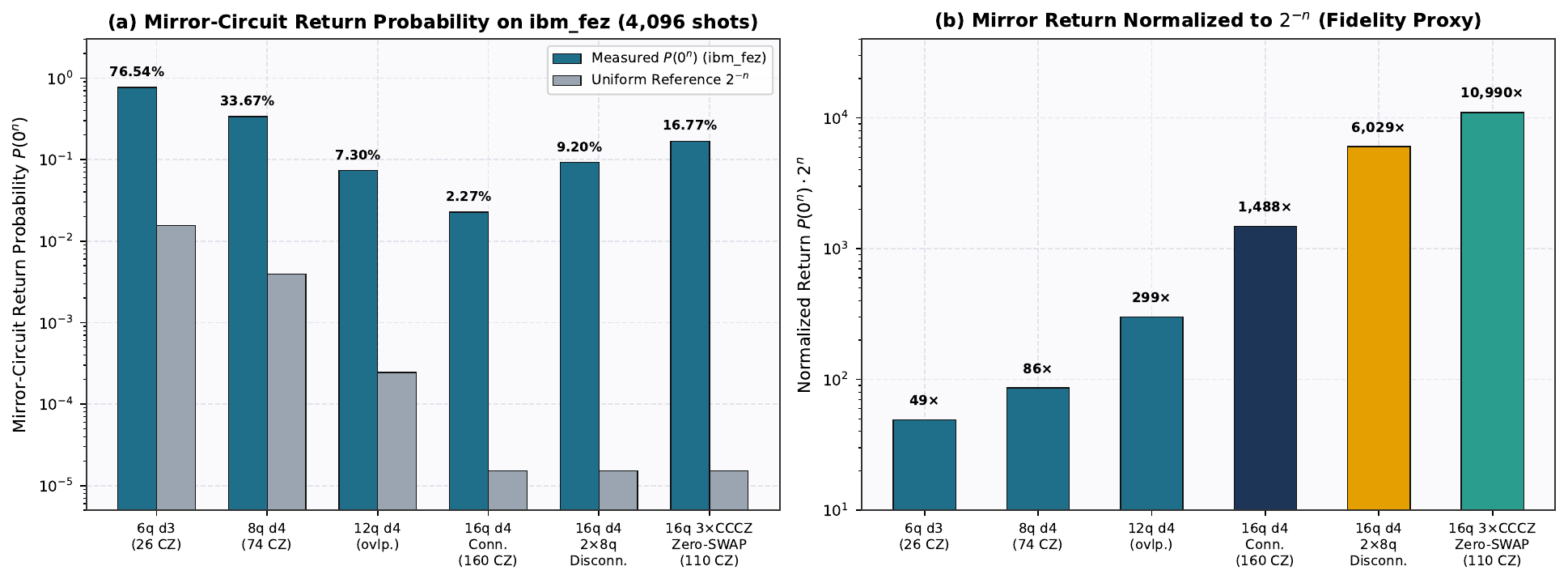}
    \caption{Mirror-circuit return probability $P(0^n)$ on \texttt{ibm\_fez} vs.\ $2^{-n}$ (a) and $P(0^n)2^n$ (b), including the 23-qubit zero-SWAP $3\times\mathrm{CCCZ}$ compilation ($16.77\%$ raw). Note that $2^{-n}$ is a uniform reference, not a noise baseline: a fully depolarized mirror circuit also returns $2^{-n}$.}
    \label{fig:fig2}
\end{figure*}

\begin{figure}[t]
    \centering
    \includegraphics[width=\columnwidth]{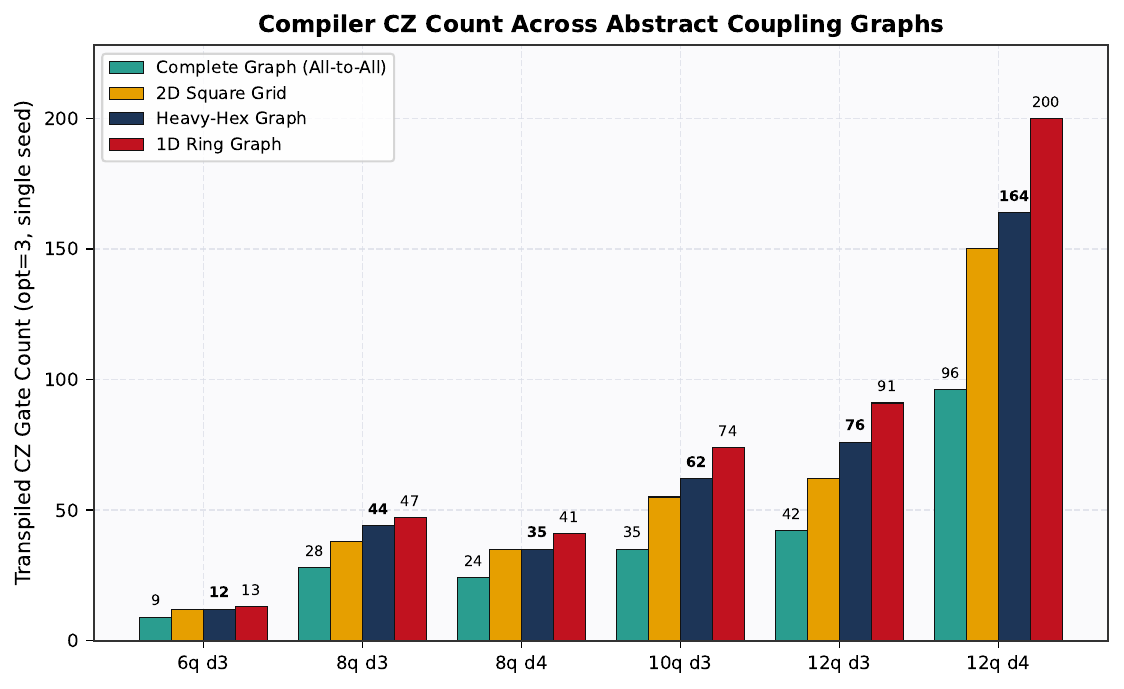}
    \caption{Transpiled CZ count across abstract coupling graphs (\texttt{opt=3}, single seed).}
    \label{fig:fig3}
\end{figure}

% =========================================================================
\section{Results}
% =========================================================================

\subsection{Flatness controls}

Table~\ref{tab:flatness} lists measured TVDs together with the ideal value and the exact expected 4,096-shot TVD of a uniform source ($0.049$, $0.099$, and $0.195$ at $n=6,8,10$).
Flat states lie near the finite-shot uniform reference (though $n=6$ RSB at $0.0735$ and $n=8$ \texttt{ibm\_fez} at $0.1147$ exceed uniform shot noise in Monte Carlo checks), and remain uninformative as coherence certificates. The non-flat controls yield nonuniformity ratios $R=\mathrm{TVD}_{\mathrm{hw}}/\mathrm{TVD}_{\mathrm{ideal}}$ of $0.90$ ($n=6$ MM), $0.70$ ($n=8$), $0.69$ ($n=10$), $0.57$ (RSB, $n=6$), and $0.89$ (mod-3 fooler, both processors). Because TVD from uniform is invariant under output-label permutations (so even $R=1$ is compatible with zero mass on the true $2^{n-2}$-outcome Walsh support of the MM controls), $R$ serves purely as a descriptive nonuniformity ratio rather than a target-coherence estimator. An early $n=7$ test (the $n=6$ MM state plus one idle $\lvert{+}\rangle$ wire, whose plug-in Walsh-peak estimator is upward-biased even noiselessly---biasing plug-in nonlinearity downward to $54.86$ at 4,096 shots, whereas depolarizing noise biases plug-in nonlinearity upward toward $\approx 58.3 > 56$) is omitted as uninformative.

\begin{table}[t]
\centering
\caption{$X$-basis TVD from uniform (4,096 shots). Ideal: exact value. Ref.: exact expected 4,096-shot TVD of an ideal uniform (or ideal non-flat) source. $R=\mathrm{TVD}_{\mathrm{hw}}/\mathrm{TVD}_{\mathrm{ideal}}$.}
\label{tab:flatness}
\resizebox{\columnwidth}{!}{%
\begin{tabular}{@{}llccccc@{}}
\toprule
State & Backend & TVD$_{\mathrm{hw}}$ & Ideal & Ref. & $R$ & $H/n$ (\%) \\
\midrule
$n=6$ d3 MM flat        & fez      & 0.0581 & 0     & 0.049 & --   & 99.74 \\
$n=6$ d3 MM control     & fez      & 0.6716 & 0.750 & 0.750 & 0.90 & 73.85 \\
$n=6$ d3 RSB flat       & fez      & 0.0735 & 0     & 0.049 & --   & 99.51 \\
$n=6$ d3 RSB control    & fez      & 0.5012 & 0.875 & 0.875 & 0.57 & 78.08 \\
$n=8$ d4 MM flat        & fez      & 0.1147 & 0     & 0.099 & --   & 99.24 \\
$n=8$ d4 MM flat        & kingston & 0.1077 & 0     & 0.099 & --   & 99.35 \\
$n=8$ d4 MM control     & fez      & 0.5256 & 0.750 & 0.750 & 0.70 & 86.14 \\
$n=8$ d4 mod-3 fooler   & fez      & 0.6235 & 0.703 & 0.703 & 0.89 & 75.31 \\
$n=8$ d4 mod-3 fooler   & kingston & 0.6243 & 0.703 & 0.703 & 0.89 & 75.46 \\
$n=10$ d3 MM flat       & fez      & 0.2007 & 0     & 0.195 & --   & 98.00 \\
$n=10$ d3 MM control    & fez      & 0.5188 & 0.750 & 0.750 & 0.69 & 88.74 \\
\bottomrule
\end{tabular}%
}
\end{table}

\subsection{Mirror circuits}

Since the $X$-basis TVD saturates at $n=16$ ($0.939$ for any near-uniform source at 4,096 shots), we ran mirror circuits $H^{\otimes n}U_f^\dagger\,|\,U_fH^{\otimes n}$ (barrier at the midpoint)~\cite{proctor2022measuring}. Ideally $P(0^n)=1$. % FIX-R3 (major)
A mirror circuit returns to $\lvert0^n\rangle$ for \emph{any} implemented unitary $V$ with $V^\dagger V=I$; it measures how well the compiled circuit composes with its inverse, not whether $U_f$ (or its non-Clifford part) was implemented correctly, and errors that are inverted by the second half are invisible. We therefore treat $P(0^n)$ as a circuit-fidelity proxy, not as a certificate of ``non-Clifford coherence''.

Results are in Table~\ref{tab:echo}. On \texttt{ibm\_fez}, the connected 16-qubit instance (antipodal orbit, two CCCZ and one $\mathrm{CCZ}(3,4,5)$ bridge; 160 CZ, depth 243) gave $93/4096$ returns ($2.27\%$). A 16-qubit chain of three overlapping CCCZ gates (230 CZ, depth 513) gave $0/4096$ (one-sided 95\% upper limit approximately $7.31\times10^{-4}$); the same instance compiled to 226 CZ (depth 502) on \texttt{ibm\_kingston} gave $9/4096$. With one point per circuit and processor, these data do not identify a decay law or establish a threshold (``cliff'').
Single runs on the two processors, executed at different times, differ in both directions: \texttt{ibm\_kingston} is $2.9\times$ better on two disconnected 8-qubit blocks ($26.39\%$ vs.\ $9.20\%$) but $4.9\times$ worse on the connected instance ($0.46\%$ vs.\ $2.27\%$). Without repeated runs we cannot attribute these differences to the processors.

In one same-job comparison of two complete compilations for the overlapping $3\times\mathrm{CCCZ}$ chain on \texttt{ibm\_fez} (job \texttt{dc3jsoqnj2vs73digg0g}, 4,096 shots/circuit), we compared the default 16-qubit compilation (222 CZ, depth 515, 2Q depth 155) against a calibration-selected 23-qubit zero-SWAP compilation using four compute/uncompute ancillas, three clean bridges, and 5-CNOT linear-nearest-neighbour relative-phase Toffoli (RCCX) decompositions~\cite{maslov2016advantages} (110 CZ, depth 119, 2Q depth 38; selected-patch mean edge CZ error $0.285\%$ among 252 enumerated candidates). Because layout selection, ancillas, gate synthesis, and depth change together, this test evaluates the bundled compilation rather than isolating routing alone. Cutting CZ count by $50.5\%$ and total depth by $76.9\%$ increased raw $P(0^{16})$ by $+16.67$~pp, from $4/4{,}096$ ($0.10\%$, 95\% CI $[0.03\%,0.25\%]$, $64\times$ random) to $687/4{,}096$ ($16.77\%$, 95\% CI $[15.64\%,17.95\%]$, $10{,}992\times$ random; ratio $171.8\times$ baseline) in the data-only circuit, and to $688/4{,}096$ ($16.80\%$) marginal, $592/4{,}096$ ($14.45\%$) unconditional joint yield (so $96/688=14.0\%$ of data returns occur with corrupted ancillas), and $592/2{,}576$ ($22.98\%$, 95\% CI $[21.37\%,24.66\%]$, $15{,}061\times$ random; ratio $235.3\times$ baseline) conditioned on clean ancilla return $P(\mathrm{anc}=0^7)=2{,}576/4{,}096=62.89\%$ (note that the 4-count baseline denominator carries $\approx 50\%$ relative uncertainty, making the absolute $+16.67$~pp gain and raw counts the primary metric rather than the wide ratio interval of roughly $67\times$--$620\times$).

\begin{table}[t]
\centering
\caption{Mirror-circuit return probability (4,096 shots). 2Q/depth from the transpiled ISA circuit (``--'' where not logged in early schema).}
\label{tab:echo}
\resizebox{\columnwidth}{!}{%
\begin{tabular}{@{}llccc@{}}
\toprule
Mirror circuit & Backend & 2Q / Depth & $P(0^n)$ & $P(0^n)2^n$ \\
\midrule
$n=6$ d3 (CCZ)                  & fez      & 26 / 80   & 76.54\% & 49 \\
$n=8$ d4 (CCCZ+CCZ)             & fez      & 74 / 230  & 33.67\% & 86.2 \\
$n=8$ d4 (CCCZ+CCZ)             & kingston & --        & 33.62\% & 86.1 \\
$n=12$ d4 (overlapping CCCZ)    & fez      & --        & 7.30\%  & 299 \\
$n=16$ d4 connected             & fez      & 160 / 243 & 2.27\%  & 1{,}488 \\
$n=16$ d4 connected             & kingston & 150 / 253 & 0.46\%  & 304 \\
$n=16$ d4 $2\times8$ disconn.   & fez      & --        & 9.20\%  & 6{,}032 \\
$n=16$ d4 $2\times8$ disconn.   & kingston & --        & 26.39\% & 17{,}296 \\
$n=16$ d4 3-CCCZ chain          & fez      & 230 / 513 & 0/4,096 (0.00\%) & $<48$ \\
$n=16$ d4 3-CCCZ chain          & kingston & 226 / 502 & 9/4,096 (0.22\%) & 144 \\
\midrule
\multicolumn{5}{@{}l}{\emph{Same-job A/B compilation comparison on \texttt{ibm\_fez} (job \texttt{dc3jsoqnj2vs73digg0g})}} \\
$n=16$ 3-CCCZ baseline (16q)    & fez      & 222 / 515 & 4/4,096 (0.10\%)   & 64 \\
$n=16$ 3-CCCZ zero-SWAP (23q)   & fez      & 110 / 119 & 687/4,096 (16.77\%) & 10{,}992 \\
\quad joint ($0^{16}\wedge\mathrm{anc}=0^7$) & fez & 110 / 119 & 592/4,096 (14.45\%) & -- \\
\quad post-sel.\ ($592/2{,}576$) & fez & 110 / 119 & 22.98\% & 15{,}061 \\
\bottomrule
\end{tabular}%
}
\end{table}

\begin{figure*}[t]
    \centering
    \includegraphics[width=0.96\textwidth]{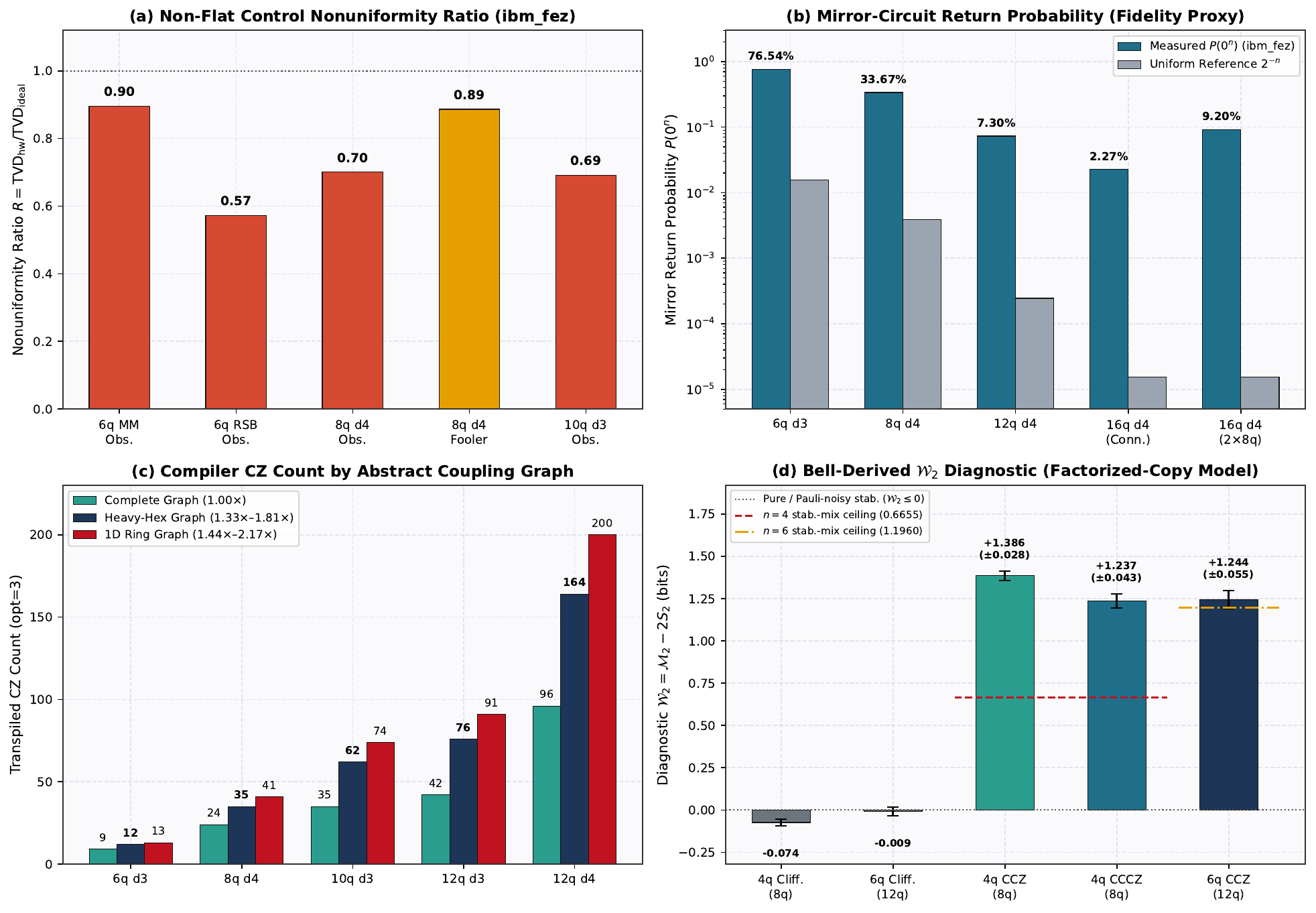}
    \caption{Executive summary on \texttt{ibm\_fez}: (a)~non-flat control nonuniformity ratio $R=\mathrm{TVD}_{\mathrm{hw}}/\mathrm{TVD}_{\mathrm{ideal}}$; (b)~mirror-circuit return probabilities $P(0^n)$ vs.\ $2^{-n}$; (c)~transpiled CZ counts across abstract coupling graphs; and (d)~Bell-derived $\mathcal{W}_2=\mathcal{M}_2-2S_2$ diagnostic under a factorized-copy, trusted-measurement model (2,000-resample bootstrap SEs) with analytic H\"older stabilizer-mixture ceilings ($0.6655\,$bits at $n=4$, $1.1960\,$bits at $n=6$).}
    \label{fig:fig4}
\end{figure*}

\subsection{Magic square and GHZ parity games}

\textbf{Magic square.} Each of the 9 row/column contexts was run with 4,096 shots (36,864 trials per configuration; Table~\ref{tab:ms_ghz}). Removing SWAPs by \texttt{opt=3} transpilation raised the raw win rate from $91.20\%$ to $96.56\%$ (job \texttt{dc3drl6q4uuc73a6osqg}). A calibration-selected chain with DD off (job \texttt{dc3e6vdhdvas73bnmbn0}) gave $96.85\pm0.09\%$, i.e.\ $+0.29$~pp ($2.2\sigma$) over \texttt{opt=3}. A run with XY4 DD on a different chain containing a qubit with $1.55\%$ readout error gave $94.45\%$; because chain and DD changed together, this run does not isolate the effect of DD. % FIX-R6 (confound)

Two calibration circuits ($\lvert0000\rangle$, $\lvert1111\rangle$) define a tensored readout model~\cite{bravyi2021mitigating}; applying this fitted correction shifts the win rate by $+1.50$~pp (to $98.35\%$, i.e.\ $47.7\%$ of the $3.15$~pp raw deficit), without separating causal readout error from gate, state-preparation, or calibration-model error.
On \texttt{ibm\_kingston} (chain 147--150) the raw and corrected values are $92.11\%$ and $98.46\%$; a $6.35$~pp correction is inconsistent with the backend-reported readout errors of these qubits ($\le0.64\%$), and we flag it as unresolved.

Google reports $0.9830\pm0.0001$ on Willow from $10^5$ shots per context~\cite{google2026contextuality}.
On heavy-hex \texttt{ibm\_fez}, our layouts depend on the column input (Alice's logical qubits are mapped to different physical qubits for $c\in\{0,1\}$ and $c=2$) and the $(r,c)=(2,2)$ context required SWAPs across the Alice/Bob partition after input-dependent gates. In a single device the classical value $8/9$ applies only to implementations in which each party's operations depend on its own input alone; our IBM numbers should be read as a fidelity benchmark of nine small Clifford circuits, with $8/9$ shown as a theoretical reference only.

\textbf{Cross-platform trapped-ion magic square (\texttt{IonQ Forte Enterprise}).} To eliminate the heavy-hex layout switching and cross-partition SWAPs, we executed all nine contexts on the 36-qubit all-to-all trapped-ion processor \texttt{IonQ Forte Enterprise} (\texttt{ionq:forte-ent} via Open Quantum; reported calibration 1Q fidelity $99.98\%$, 2Q fidelity $98.63\%$, $T_1=188\,$s, $T_2=950\,$ms) using three parallel 12-qubit jobs (one job per row, packing columns $c=0,1,2$ onto disjoint 4-qubit registers, $160$ shots/job $= 1{,}440$ total plays; Table~\ref{tab:jobs}). Every context uses a \emph{fixed} Alice/Bob logical qubit partition with $2$ Bell-pair entangling CNOTs followed strictly by Alice-only and/or Bob-only local gates ($2$--$4$ logical CNOTs/play, mean $2.67$, zero cross-partition SWAPs; native compiled trapped-ion pulse counts are not exposed by the submission endpoint). Without readout mitigation, we obtain $1{,}401/1{,}440 = \mathbf{97.29\%}$ (Row~0 $96.67\%$, Row~1 $99.17\%$, Row~2 $96.04\%$). Accounting for mild positive within-shot correlation across the three parallel 4-qubit blocks sharing the same 12-qubit ion chain (pairwise $r$ up to $+0.147$ in Row~2), the cluster-robust per-shot standard error across the 480 12-qubit shots is $\pm0.45\%$ ($95\%$ CI $[96.42\%, 98.17\%]$), compared with $\pm0.43\%$ ($[96.46\%, 98.13\%]$) under an independent-play model. Relative to our best raw IBM Heron~r2 chain ($96.85\pm0.09\%$), the IonQ--IBM difference for these selected jobs is $+0.45$~pp (95\% shot-model CI $[-0.45,+1.34]$~pp; run-to-run, day-to-day, and calibration variability lie outside this interval), and compares favourably with the rough mean 2Q gate-survival heuristic $(4\cdot0.9863^2+4\cdot0.9863^3+0.9863^4)/9=96.39\%$ (not a hard ceiling, because not every 2Q error flips the checked parity, as illustrated by Row~1 reaching $99.17\%$). Crucially, however, packing column $c$ into a fixed physical block ($c=0,1,2$ on registers $0$--$3$, $4$--$7$, $8$--$11$) reveals the column identity to any block-dependent classical strategy: Bob can fix a valid triple per column, and Alice---knowing her row and block index---copies the intersection sign and completes her row parity, winning $9/9$ ($100\%$) without communication. This $100\%$ block-specific classical ceiling persists even if the three blocks were perfectly isolated (independently of the shared motional phonon bus and static batch compilation); therefore, $8/9$ remains strictly a theoretical reference under column-blind Alice responses, while $97.29\pm0.45\%$ stands as a clean circuit-fidelity benchmark of the nine fixed-partition circuits.

\textbf{GHZ parity circuits.} For $n\in\{6,10,16,20,30\}$ we drew 8 even-weight inputs per size (seed 2026; 1,024 shots each) and prepared GHZ states on calibration-selected spanning trees with 0 SWAPs (job \texttt{dc3fqemq4uuc73a6s6cg}; Table~\ref{tab:ms_ghz}). Between-input homogeneity is rejected by a Pearson test at $n=6$ ($p=6.5\times10^{-4}$) and $n=10$ ($p=0.017$). Crucially, as noted in Sec.~\ref{sec:games}, the exact deterministic local ceiling on our 8 fixed inputs is $87.5\%$ at $n=6,10$ and $100\%$ at $n=16,20,30$ (and $100\%$ for the 2-input $n=50$ test); thus only $n=6$ ($89.44\%$) exceeds its restricted local ceiling, and we omit fixed-question $z$ scores against the uniform-input value.

Across four earlier $n=30$ jobs with nominally equivalent 29-CZ trees (different inputs and dates), win rates ranged from $47.47\%$ to $59.46\%$ (descriptive mean $53.85\%$, s.d.\ $5.06$~pp). Values below $1/2$ are consistent with (though do not uniquely identify) a collective phase offset $\phi_0$. To probe parity oscillations directly, we ran an 8-angle global-phase scan ($\phi \in \{0,\pi/4,\dots,7\pi/4\}$ via $R_Z(\phi/30)$ on all 30 qubits) alongside three same-job all-$X$ parity checks of linear, binary-tree, and 0-SWAP BFS trees with XY4 DD (job \texttt{dc3j25lhdvas73bo0v3g}, 2,048 shots/circuit). An unweighted harmonic fit $\langle X^{\otimes 30}\rangle(\phi) = V\cos(\phi+\phi_0)$ gives a fitted parity harmonic $V=0.1560\pm0.0110$ and phase offset $\phi_0=-70.9^\circ\pm4.0^\circ$ (nominal fitted peak $57.80\%$; zero-offset $\chi^2=12.41$ on 6 df; a test of zero sine and cosine components has 2 df). Because eight angles over one nominal cycle do not project out complementary coherences of Hamming weight $w=1,\dots,29$, we label $V$ a descriptive fitted parity harmonic rather than isolated GHZ coherence $\lvert\rho_{0^{30},1^{30}}\rvert$~\cite{mooney2021generation,javadiabhari2025bigcats}. The three same-job all-$X$ parity checks give $62.42\%$ (s.d.\ $0.64$~pp) for linear, $62.00\%$ (s.d.\ $1.56$~pp) for binary tree, and $53.96\%$ (s.d.\ $0.98$~pp) for 0-SWAP BFS (within-circuit shot SE $\approx 1.1$~pp).

\textbf{DD ablation at $n=30$.} In two 24-circuit jobs (DD off: \texttt{dc3hu6dhdvas73bnucbg}; XY4: \texttt{dc3i0i6q4uuc73a71gfg}) using the same 8 inputs (drawn with a different generator from the main run), XY4 raised a linear chain from $41.28\%$ to $57.04\%$ and the 0-SWAP tree from $47.47\%$ to $52.73\%$, and lowered a SWAP-routed binary tree from $57.36\%$ to $55.21\%$. Because DD off/on were separate jobs measured only at fixed phase $\phi=0$ (without a phase sweep), both time drift and coherent phase shifts $\Delta\phi_0$ are confounded with changes in visibility $V$ (which also explains why the 0-SWAP BFS tree, with $\phi_0=-70.9^\circ$, trails linear and binary trees at $\phi=0$); the data are consistent with, but do not demonstrate, suppression of $ZZ$-induced phase errors by DD~\cite{tripathi2022suppression,ezzell2023dynamical}.

For context, GHZ games have been won above the classical value up to $N=71$ qubits on Willow~\cite{google2026contextuality}, and IBM has reported GHZ fidelity $0.56(3)$ at 120 qubits conditional on $28\%$ post-selection~\cite{javadiabhari2025bigcats}; our GHZ data are a benchmark of these specific circuits, not a scaling record.

\begin{table}[t]
\centering
\caption{Magic square (\texttt{ibm\_fez} and \texttt{IonQ Forte Enterprise}) and GHZ parity circuits (\texttt{ibm\_fez}); raw win rates with binomial SE (cluster-robust per-shot SE for parallel IonQ jobs). Uniform-input classical values are theoretical references only; for the fixed GHZ inputs sampled, the exact restricted local ceiling is $7/8$ ($n=6,10$), $8/8$ ($n=16,20,30$), and $2/2$ (2-input $n=50$).}
\label{tab:ms_ghz}
\resizebox{\columnwidth}{!}{%
\begin{tabular}{@{}lccc@{}}
\toprule
Configuration & Win rate ($\pm\sigma$) & Uniform ref. & Restricted ceiling \\
\midrule
\multicolumn{4}{@{}l}{\emph{Magic square (IBM \texttt{ibm\_fez}: 36,864 trials; IonQ Forte Ent.: 1,440 plays / 480 shots)}} \\
\texttt{opt=0}, 7.33 CZ (\texttt{ibm\_fez})  & $91.20 \pm 0.15\%$ & 88.89\% & -- \\
XY4 DD, chain 21--24 (\texttt{ibm\_fez})   & $94.45 \pm 0.12\%$ & 88.89\% & -- \\
\texttt{opt=3}, 3.00 CZ (\texttt{ibm\_fez})  & $96.56 \pm 0.09\%$ & 88.89\% & -- \\
Chain 123/136/143/144, DD off (\texttt{ibm\_fez}) & $96.85 \pm 0.09\%$ & 88.89\% & -- \\
\quad readout-corrected$^\dagger$       & 98.35\%            & --      & -- \\
\textbf{\texttt{IonQ Forte Ent.} (all-to-all, 2.67 CX)} & $\mathbf{97.29 \pm 0.45\%}$ & 88.89\% & 100.0\% (9/9)$^\ast$ \\
\midrule
\multicolumn{4}{@{}l}{\emph{GHZ circuits, 8 fixed inputs/$n$, 8,192 shots/$n$ (job \texttt{dc3fqemq4uuc73a6s6cg})}} \\
$n=6$ (5 CZ, depth 15)   & $89.44 \pm 0.34\%$ & 62.50\% & 87.50\% (7/8) \\
$n=10$ (9 CZ, depth 21)  & $84.44 \pm 0.40\%$ & 53.13\% & 87.50\% (7/8) \\
$n=16$ (15 CZ, depth 27) & $76.20 \pm 0.47\%$ & 50.39\% & 100.0\% (8/8) \\
$n=20$ (19 CZ, depth 29) & $72.61 \pm 0.49\%$ & 50.10\% & 100.0\% (8/8) \\
$n=30$ (29 CZ, depth 36) & $59.46 \pm 0.54\%$ & 50.00\% & 100.0\% (8/8)$^\ddagger$ \\
\midrule
\multicolumn{4}{@{}l}{\emph{Two fixed inputs, 8,192 shots (job \texttt{dc3e2ouq4uuc73a6p6a0})}} \\
$n=50$ (49 CZ, 2 inputs) & $47.44 \pm 0.55\%$ & 50.00\% & 100.0\% (2/2) \\
\bottomrule
\end{tabular}%
}
\vspace{2pt}
\parbox{\columnwidth}{\scriptsize $^\dagger$Fitted two-calibration tensored model ($\lvert0000\rangle$/$\lvert1111\rangle$), simplex-projected ($98.38\%$ unconstrained). $^\ast$Block position encodes column $c$ in the parallel 12q layout, yielding a $100\%$ ($9/9$) classical ceiling under block-dependent strategies. $^\ddagger$Earlier $n=30$ runs across different jobs/inputs range from $47.5\%$ to $59.5\%$ (text).}
\end{table}

\subsection{Non-Clifford tests}

\textbf{CCZ games.} On $\mathrm{CCZ}\lvert{+}{+}{+}\rangle$ (job \texttt{dc3e8othdvas73bnme30}, 4,096 shots per context; transpiled contexts use $7$--$9$ CZ gates depending on basis rotation), the Mermin-type game gives $M=+2.716$ and $13{,}754/16{,}384=83.95\pm0.29\%$ (ideal $87.99\%$), and the Hardy-type game gives $96.65\pm0.16\%$ on the three nonlinear contexts, $18.65\%$ on $XXX$ (ideal $18.75\%$), and $12{,}640/16{,}384=77.15\%$ overall (fixed-context stratified shot SE $0.20$~pp; pooled SE $0.33$~pp; ideal $79.69\%$), versus the $75\%$ theoretical local-strategy reference. Under an idealized 16,384-trial independent-shot local-strategy model (non-adaptive fixed contexts), a one-sided Hoeffding bound gives $p\le2.7\times10^{-7}$ for the Hardy margin (or $p\le 2.3\times10^{-2}$ if shots are conservatively grouped into 4,096 four-context tuples); however, these are single-chip circuits with context-dependent transpilation known at compile time, closing neither locality nor freedom-of-choice loopholes and standing alongside prior photonic hypergraph-nonlocality demonstrations~\cite{huang2024hypergraph}.

\textbf{Target-generator checks and 16-generator test.} Table~\ref{tab:nonclifford} lists the single-context checks defined in Sec.~\ref{sec:games}. The earlier $n=16$ two-check job (\texttt{dc3e8othdvas73bnme30}) used a $\mathrm{CCZ}(2,3,4)$ bridge and dropped both $\mathrm{CZ}(0,8)$ and $\mathrm{CZ}(7,15)$ in its control; its two checks are disjoint 5-qubit tests blind to the bridge (covariance-aware SEs on the two-check mean: $0.47$~pp flat, $0.56$~pp control). Separately, on the $\mathrm{CCZ}(3,4,5)$-bridged 16-qubit degree-4 circuit matching the connected mirror experiment, we measured all 16 target generators $K_0,\dots,K_{15}$ in job \texttt{dc3j25lhdvas73bo0v3g} (9 separately transpiled circuits: 1 joint circuit for $K_8$--$K_{15}$ and 8 circuits for $K_0$--$K_7$; avg.\ $84$ CZ, depth $191$, 2,048 shots/circuit). The $Z$-register generators $K_8$--$K_{15}$ average $85.32\%$, the $Y$-register generators $K_0$--$K_7$ average $83.45\%$, and the three target-generator tests whose scoring includes the $\mathrm{CCZ}(3,4,5)$ bridge terms (supports of 7, 6, and 6 qubits) yield $K_3=1{,}666/2{,}048$ ($81.35\%$), $K_4=1{,}723/2{,}048$ ($84.13\%$), and $K_5=1{,}750/2{,}048$ ($85.45\%$), averaging $83.64\%$. The overall 16-generator mean is $84.39\%$, below the $15/16=93.75\%$ mean threshold required for a positive union-bound fidelity lower bound~\cite{morimae2017verification,zhu2019efficient}; the unreduced plug-in union bound is $-1.50$, so the physical global fidelity lower bound is trivially $0$ (and a global verification bound would additionally require a common physical preparation across contexts). Crucially, even assuming a common state preparation and ideal measurements, the 16 published generator marginals alone do \emph{not} certify entanglement across the $\mathrm{CCZ}(3,4,5)$ bridge: removing $\mathrm{CCZ}(3,4,5)$ while keeping all internal phase gates of each 8-qubit half, preparing $(q_4,q_5)$ in $\sqrt{0.50}\lvert00\rangle+\sqrt{0.26}\lvert01\rangle+\sqrt{0.24}\lvert10\rangle$ (zero $\lvert11\rangle$ amplitude, strictly separable across the half-cut) and all remaining qubits in $\lvert{+}\rangle$ yields exact pass probabilities $p_3=100.00\%$, $p_4=84.64\%$, $p_5=86.06\%$, and $p_j=100.00\%$ for the other 13 generators. Because a local $Z_j$ error flips exclusively $K_j$, applying independent $Z_j$ flips with probabilities $(p_j-q_j)/(2p_j-1)\in[0.74\%,19.87\%]$ reproduces all 16 archived IBM pass rates $q_j$ to machine precision ($<2\times10^{-16}$) without any cross-bridge entanglement; moreover, because the ideal state starts with all eight $Z$-wing signs $K_8,\dots,K_{15}$ equal to $+1$, drawing a correlated pattern of local $Z_8,\dots,Z_{15}$ flips likewise reproduces any target joint distribution over $K_8$--$K_{15}$ while preserving strict separability across the bridge.

\begin{table}[t]
\centering
\caption{CCZ games and single-context target-generator checks on \texttt{ibm\_fez} (raw; fixed-context stratified shot SE for the 4-context games, covariance-aware SE for the $n=16$ two-check mean).}
\label{tab:nonclifford}
\resizebox{\columnwidth}{!}{%
\begin{tabular}{@{}lccc@{}}
\toprule
Test & Measured & Reference & Ideal \\
\midrule
\multicolumn{4}{@{}l}{\emph{CCZ games (theoretical local-strategy ref.\ 75\%; 16,384 shots)}} \\
Mermin-type, $M=+2.716$              & $83.95 \pm 0.29\%$ & 75.00\% & 87.99\% \\
Hardy-type (4 contexts)              & $77.15 \pm 0.20\%$ & 75.00\% & 79.69\% \\
\quad 3 nonlinear contexts           & $96.65 \pm 0.16\%$ & --      & 100\% \\
\midrule
\multicolumn{4}{@{}l}{\emph{Target-generator checks: flat vs.\ obstructed control (antipodal CZ removed/replaced)}} \\
$n=6$ d3 (4-qubit check)             & $92.61 \pm 0.24\%$ & $50.72 \pm 0.45\%$ & 100 / 50\% \\
$n=8$ d4 (5-qubit check)             & $93.68 \pm 0.38\%$ & $50.88 \pm 0.78\%$ & 100 / 50\% \\
$n=12$ CCZ+CCCZ (5-qubit check)      & $94.12 \pm 0.37\%$ & $48.24 \pm 0.78\%$ & 100 / 50\% \\
$n=16$ $\mathrm{CCZ}(2,3,4)$, 1-check mean$^\ast$ & $78.02 \pm 0.47\%$ & $50.46 \pm 0.56\%$ & 100 / 50\% \\
$n=16$ $\mathrm{CCZ}(2,3,4)$, both checks$^\ast$  & $61.89 \pm 0.76\%$ & $26.10 \pm 0.69\%$ & 100 / 25\% \\
\bottomrule
\end{tabular}%
}
\vspace{2pt}
\parbox{\columnwidth}{\scriptsize $^\ast$Control drops both $\mathrm{CZ}(0,8)$ and $\mathrm{CZ}(7,15)$; distinct from the $\mathrm{CCZ}(3,4,5)$ circuit used in the 16-generator test (\texttt{dc3j25lhdvas73bo0v3g}).}
\end{table}

\begin{figure*}[t]
    \centering
    \includegraphics[width=0.98\textwidth]{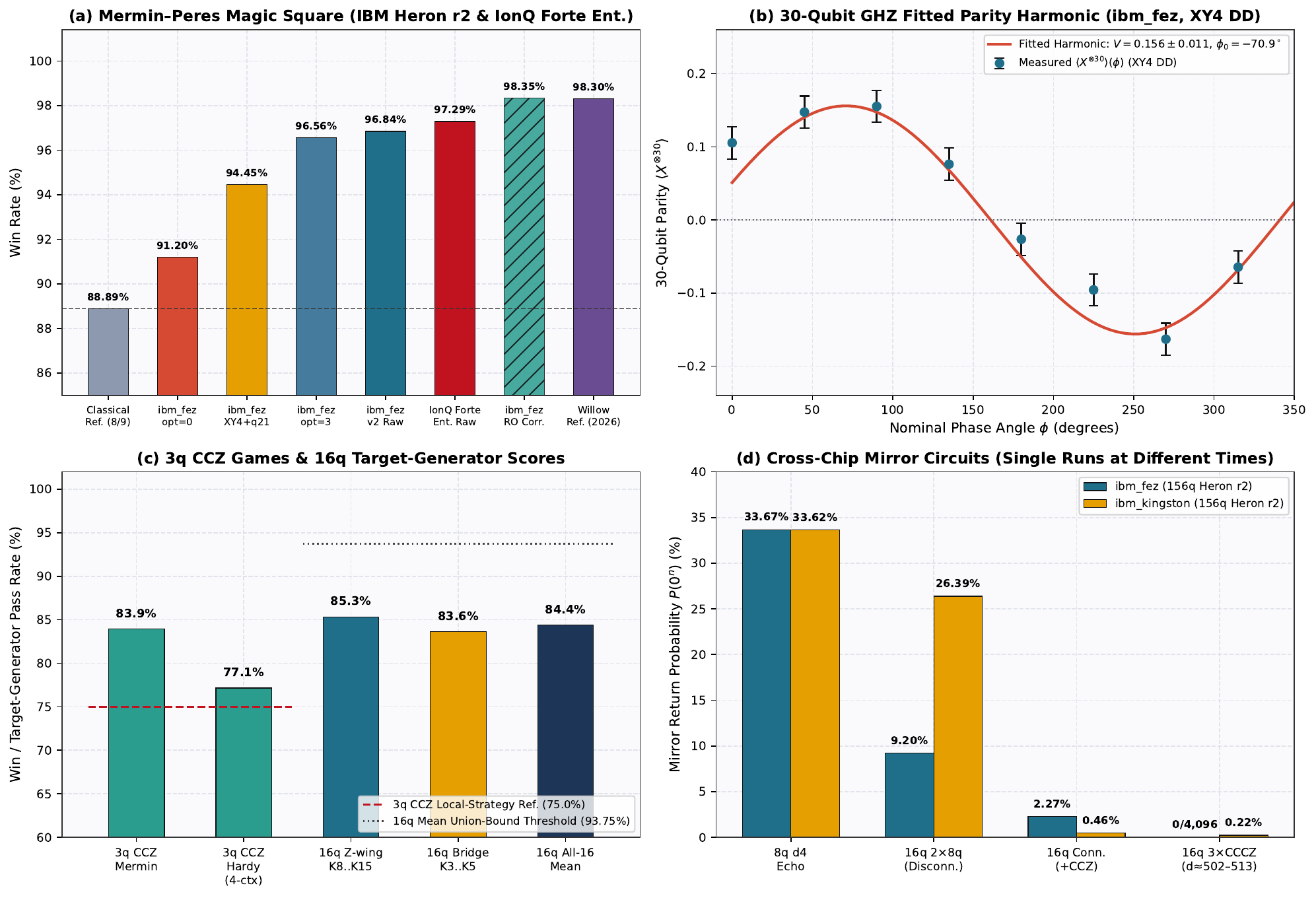}
    \caption{(a)~Magic-square configurations on \texttt{ibm\_fez} and all-to-all \texttt{IonQ Forte Enterprise} ($97.29\%$ raw, fixed partition; hatched: fitted IBM readout correction; Willow reference from~\cite{google2026contextuality}). (b)~30-qubit GHZ fitted parity harmonic with XY4 DD (job \texttt{dc3j25lhdvas73bo0v3g}, $V=0.156\pm0.011$, $\phi_0=-70.9^\circ$). (c)~3-qubit CCZ games vs.\ $75\%$ local-strategy reference and 16-generator scores vs.\ the $93.75\%$ mean union-bound threshold. (d)~Cross-processor mirror-circuit comparison on \texttt{ibm\_fez} and \texttt{ibm\_kingston} (single runs at different times; zero count labeled \texttt{0/4,096}).}
    \label{fig:fig5}
\end{figure*}

\subsection{Bell-derived magic diagnostic from two-copy sampling}

Two copies of $\lvert\psi_f\rangle$ were prepared on $2n$ qubits and measured in the Bell basis ($\mathrm{CX}_{i,i+n}$, then $H_i$) with 8,192 shots (job \texttt{dc3i8rqnj2vs73didjo0})~\cite{montanaro2017bell,gross2021schur,haug2023scalable}. Under an ideal Bell measurement on two identical copies $\rho^{\otimes 2}$, the unbiased collision U-statistic $\hat C$ gives $\mathcal{M}_{2}=-\log_2(2^n\hat C)$, which equals the pure-state stabilizer R\'enyi entropy~\cite{leone2022stabilizer} for pure states and $-\log_2\!\big(2^{-n}\sum_P\mathrm{tr}(\rho P)^4\big)$ for mixed states, while the parity of $\lvert11\rangle$ pairs estimates $\mathrm{tr}(\rho^2)$ and $S_2=-\log_2\mathrm{tr}(\rho^2)$~\cite{garcia2013swap,islam2015measuring,huang2022quantum}.

We report two Bell-derived quantities with 2,000-resample multinomial bootstrap standard errors (Table~\ref{tab:magic}):
(i)~the mixed-state SRE $\widetilde{\mathcal{M}}_2=\mathcal{M}_2-S_2$~\cite{leone2022stabilizer}, which is positive for the noisy Clifford controls ($0.298\pm0.012$ and $0.469\pm0.015\,$bits); and (ii)~the purity-normalized Bell diagnostic $\mathcal{W}_2=\mathcal{M}_2-2S_2$~\cite{haug2023scalable,haug2025witnessing}. For any \emph{pure} or \emph{Pauli-noisy} stabilizer state under ideal Bell measurements, $\mathcal{W}_2 \le 0$. Crucially, however, as analyzed in detail in our companion study~\cite{vono2026ccz832}, $\mathcal{W}_2$ is \emph{not} bounded by $0$ on general convex mixtures of non-commuting stabilizer states: for any factorized product $\rho\otimes\sigma$ of $n$-qubit stabilizer mixtures ($L=2^n-1$), Cauchy--Schwarz and H\"older's inequality yield the analytic upper bound $2^{\mathcal{W}_2} \le \max_{s\in[0,L]} \frac{(1+s)^2}{2^n(1+s^3/L^2)}$, giving conservative ceilings of $\mathcal{W}_2 < 0.6655\,$bits at $n=4$ ($L=15$; $0.6514$ achieved by $\frac{1}{2}(\lvert0^4\rangle\langle0^4\rvert+\lvert+^4\rangle\langle+^4\rvert)$) and $\mathcal{W}_2 < 1.1960\,$bits at $n=6$ ($L=63$; $1.1819$ achieved by a $3$-MUB stabilizer mixture). We obtain $\mathcal{W}_2=-0.074\pm0.019$ and $-0.009\pm0.024\,$bits for the Clifford controls, versus $+1.386\pm0.028$ (descriptive bootstrap-SE distance $+26.2$ above $0.6655$ from unrounded SE $0.0275$) for $n=4$ CCZ, $+1.237\pm0.043$ ($+13.4$ above $0.6655$, unrounded SE $0.0427$) for $n=4$ CCCZ, and $+1.245\pm0.055\,$bits for $n=6$ flat CCZ (only $+0.89$ bootstrap SE above the global $n=6$ ceiling $1.1960$, hence inconclusive under the global bound; note that conditioning the H\"older bound on the measured swap expectation $\gamma$ via $\Pi_{\mathrm{Bell}} \ge 2^{-n}[1+(2^n\gamma-1)_+^3/(2^n-1)^2]$ tightens the pointwise $n=6$ bound at $\hat{\gamma}=0.4814$ to $0.9506\,$bits, warranting a joint shot-level test on the raw histogram in future work). These SE ratios are purely descriptive distances in multinomial bootstrap SEs, not calibrated null-hypothesis $z$-scores or systematic measurement-error bounds.
Three caveats are essential. First, the two copies occupy different physical qubits ($\rho_A\otimes\rho_B$), so SWAP estimates the empirical swap-test expectation $\hat{\gamma}=\mathrm{tr}(\rho_A\rho_B)$ rather than $\mathrm{tr}(\rho^2)$; moreover, without copy factorization, separable correlated mixtures of stabilizer states can reproduce the entire ideal Bell histogram~\cite{vono2026ccz832}, and uncharacterized Bell-measurement POVM errors can bias both $S_2$ and $\mathcal{M}_2$. We therefore treat $\mathcal{W}_2$ as a \emph{Bell-derived diagnostic under a factorized-copy, trusted-measurement model}, not an unconditioned physical certification. Second, in a multinomial bootstrap around an empirical distribution, the expected resample collision rate is shifted upward by $(1-\hat C)/S$ ($\approx 8\%$ for the $n=6$ CCZ state, shifting its percentile interval below the point estimate); we therefore report bootstrap SEs descriptively rather than claiming calibrated null-tail significances. Third, $\mathcal{W}_2$ is not an estimate of pure-state magic: under global depolarizing noise at purity $0.62$, $\mathcal{W}_2$ drops to $1.247$~bits for the $n=4$ CCZ state (and $0.979$~bits for a bare $n=3$ CCZ state).

\begin{table}[t]
\centering
\caption{Two-copy Bell sampling on \texttt{ibm\_fez} (8,192 shots, 2,000-resample multinomial bootstrap SEs from archived JSON). $\hat{\gamma}$: empirical swap-test expectation ($\mathrm{tr}\rho^2$ under $\rho^{\otimes 2}$); $\mathcal{M}_2(\psi)$: exact pure-state value; $\widetilde{\mathcal{M}}_2=\mathcal{M}_2-S_2$~\cite{leone2022stabilizer}; $\mathcal{W}_2=\mathcal{M}_2-2S_2$~\cite{haug2023scalable,haug2025witnessing} compared against the $n$-qubit H\"older stabilizer-mixture ceiling $B_n$ ($0.6655$ at $n=4$, $1.1960$ at $n=6$ under $\rho\otimes\sigma$); parenthesized ratios $\Delta/\mathrm{SE}_{\mathrm{boot}}$ are descriptive bootstrap-SE distances, not calibrated null $z$-scores. All in bits.}
\label{tab:magic}
\resizebox{\columnwidth}{!}{%
\begin{tabular}{@{}lcccccc@{}}
\toprule
State ($2n$ qubits) & $\hat{\gamma}$ & $S_2$ & $\mathcal{M}_2$ & $\widetilde{\mathcal{M}}_2$ & $\mathcal{W}_2$ ($\Delta/\mathrm{SE}_{\mathrm{boot}}$) & $\mathcal{M}_2(\psi)$ \\
\midrule
$n=4$ Clifford (8q)        & $0.773(7)$  & $0.372(13)$ & $0.670(16)$ & $0.298(12)$ & $-0.074(19)$          & 0 \\
$n=6$ Clifford (12q)       & $0.719(8)$  & $0.477(16)$ & $0.945(20)$ & $0.469(15)$ & $-0.009(24)$          & 0 \\
$n=4$ CCZ (8q)             & $0.739(7)$  & $0.437(14)$ & $2.261(24)$ & $1.823(21)$ & $\mathbf{1.386(28)}$ ($+26.2$) & 1.541 \\
$n=4$ CCCZ (8q)            & $0.567(9)$  & $0.820(24)$ & $2.876(27)$ & $2.057(26)$ & $\mathbf{1.237(43)}$ ($+13.4$) & 1.378 \\
$n=6$ flat CCZ (12q)       & $0.481(10)$ & $1.055(29)$ & $3.354(37)$ & $2.299(36)$ & $\mathbf{1.245(55)}$ ($+0.89$) & 1.541 \\
\bottomrule
\end{tabular}%
}
\end{table}

% =========================================================================
\section{Discussion and Limitations}
% =========================================================================

\begin{enumerate}
    \item \textbf{What the data establish.} Our sharpest controlled finding is the same-job A/B compilation comparison on \texttt{ibm\_fez}, where a calibrated 23-qubit zero-SWAP ancilla+RCCX synthesis raises $3\times\mathrm{CCCZ}$ mirror return by $+16.67$~pp raw ($4/4{,}096$ to $687/4{,}096$; $22.98\%$ post-selected, $14.45\%$ unconditional joint yield). Non-flat controls yield descriptive nonuniformity ratios $R=0.57$--$0.90$ at $n\le10$; mirror circuits return above $2^{-n}$ up to $\sim$230 CZ; the magic-square circuits reach $96.85\pm0.09\%$ raw on \texttt{ibm\_fez} and $97.29\pm0.45\%$ raw (cluster-robust per-shot SE; Row~1 $99.17\%$; IonQ--IBM difference $+0.45$~pp, 95\% shot-model CI $[-0.45,+1.34]$~pp) on all-to-all \texttt{IonQ Forte Enterprise}; the 30-qubit GHZ scan exhibits a fitted parity harmonic $V=0.1560\pm0.0110$; the CCZ Mermin- and Hardy-type games exceed the $75\%$ theoretical local-strategy reference; 16 target-generator tests on the $\mathrm{CCZ}(3,4,5)$ circuit average $84.39\%$ ($83.64\%$ on the three bridge-sensitive generators); and under a factorized-copy, trusted-measurement model, $\mathcal{W}_2$ is negative/zero on Clifford controls while $n=4$ CCZ/CCCZ states ($+1.386, +1.237\,$bits) exceed the $n=4$ H\"older stabilizer-mixture ceiling ($0.6655\,$bits).
    \item \textbf{What they do not establish.} Flatness of bent states does not certify coherence, and $R$ is invariant under output-label permutations (so even $R=1$ is compatible with zero mass on the true Walsh support); mirror circuits do not verify the non-Clifford part; single-device games close no locality or freedom-of-choice loophole, assigning columns to fixed blocks in our parallel 12-qubit IonQ layout yields a $100\%$ ($9/9$) block-dependent classical ceiling, and our 8-input GHZ runs do not beat their restricted local ceilings for $n\ge10$; the 8-angle $n=30$ scan does not isolate the target GHZ coherence $\lvert\rho_{0^{30},1^{30}}\rvert$; the 16-generator union bound is trivially zero and the 16 archived marginals admit an explicit half-separable counterexample with local $Z_j$ flips ($0.74\%$--$19.87\%$) and zero cross-bridge entanglement; and $\mathcal{W}_2$ is inconclusive at $n=6$ ($+0.89$ bootstrap SE above $1.1960$) and is not an unconditioned hardware certification without copy factorization and bounded Bell-POVM errors (see our companion analysis~\cite{vono2026ccz832}).
    \item \textbf{Outlook.} Matched multi-placement synthesis benchmarks, Walsh-support mass metrics, Fourier-isolated GHZ coherence, randomized-block magic square layouts, and encoded logical $\mathrm{CCZ}$ error-detection vs.\ ZNE benchmarks with frozen-preparation single-block certification across superconducting and trapped-ion processors (pursued in our companion study~\cite{vono2026ccz832}) are natural next steps.
\end{enumerate}

% =========================================================================
\section*{Acknowledgments}
% =========================================================================
We acknowledge the use of IBM Quantum services (IBM Quantum Open Plan) and Open Quantum (free Spark credits on IonQ Forte Enterprise) for this work. The views expressed are those of the authors and do not reflect the official policy or position of IBM, IonQ, or Open Quantum. The authors received no specific financial funding for this work.

\section*{Author contributions, competing interests}
F.M.V.: Conceptualization, Methodology, Software, Investigation, Formal analysis, Visualization, Writing -- original draft, Writing -- review \& editing. V.G.S.\ and A.O.N.T.: Validation, Formal analysis, Writing -- review \& editing. The authors declare no competing interests.

\section*{Data and Code Availability}
All 17 IBM summary JSONs, the 3 raw 12-qubit IonQ Forte Enterprise bitstring histograms and their exact OpenQASM~2.0 circuits (\texttt{ionq\_ms\_row\{0,1,2\}\_12q.qasm}), Python experiment and analysis scripts, a pure-Python offline verification and dynamic figure-replay script (\texttt{replay\_all\_tables\_and\_figures.py}, requiring only NumPy and Matplotlib), and LaTeX/figure sources are archived on Figshare~\cite{vono2026archive} (DOI \texttt{10.6084/m9.figshare.35120473}). Offline replay independently recounts all 1,440 IonQ magic-square plays, verifies the 16-generator separable counterexample, and dynamically regenerates Figs.~1--5 from the archived JSONs; full per-shot IBM bitstring histograms (including those underlying the 2,000-resample Bell bootstrap) are retained on IBM Quantum under the 17 IBM job IDs in Table~\ref{tab:jobs} (3.21~RU account-balance difference).

\begin{table*}[t]
\centering
\caption{All 20 auditable hardware jobs across 3 QPUs: 17 on IBM Heron~r2 (4,096 shots/circuit unless noted; GHZ random/ablation: 1,024 shots/input; Bell sampling: 8,192 shots/circuit; phase scan \& 16-stabilizer verification: 2,048 shots/circuit) and 3 parallel 12-qubit jobs on \texttt{IonQ Forte Enterprise} (160 shots/job $=1{,}440$ magic-square plays).}
\label{tab:jobs}
\footnotesize\setlength{\tabcolsep}{4pt}
\resizebox{\textwidth}{!}{%
\begin{tabular}{@{}lll@{}}
\toprule
Job ID & Backend & Experiment \\
\midrule
\texttt{dc2s3ainj2vs73dh8udg} & \texttt{ibm\_fez}      & $n=6$ d3 MM flat vs.\ control \\
\texttt{dc2s6manj2vs73dh94m0} & \texttt{ibm\_fez}      & $n=8,10$ flatness; 6q/8q mirror circuits \\
\texttt{dc2sbsls2omc738chju0} & \texttt{ibm\_fez}      & RSB 6q, mod-3 fooler, 16q flatness, 12q/16q ($2\times8$) mirror \\
\texttt{dc2tabds2omc738cj42g} & \texttt{ibm\_fez}      & Connected 16q and 3-CCCZ mirror circuits \\
\texttt{dc2sc1ts2omc738chk40} & \texttt{ibm\_kingston} & 8q flatness, fooler; 8q and 16q ($2\times8$) mirror \\
\texttt{dc3drl6q4uuc73a6osqg} & \texttt{ibm\_fez}      & Magic square, \texttt{opt=0} vs.\ \texttt{opt=3} \\
\texttt{dc3dus5hdvas73bnlu6g} & \texttt{ibm\_fez}      & Magic square with XY4 DD; 6q stabilizer check \\
\texttt{dc3e2ouq4uuc73a6p6a0} & \texttt{ibm\_fez}      & GHZ game 6q--50q (2 inputs); 8q/12q stabilizer checks \\
\texttt{dc3e6vdhdvas73bnmbn0} & \texttt{ibm\_fez}      & Magic square, selected chain + readout calibration \\
\texttt{dc3e8othdvas73bnme30} & \texttt{ibm\_fez}      & CCZ games; 16q stabilizer checks \\
\texttt{dc3fqemq4uuc73a6s6cg} & \texttt{ibm\_fez}      & GHZ game 6q--30q, 8 inputs/$n$ \\
\texttt{dc3hu6dhdvas73bnucbg} & \texttt{ibm\_fez}      & 30q GHZ ablation, DD off \\
\texttt{dc3i0i6q4uuc73a71gfg} & \texttt{ibm\_fez}      & 30q GHZ ablation, XY4 DD \\
\texttt{dc3i38mq4uuc73a71mi0} & \texttt{ibm\_kingston} & Magic square; connected 16q and 3-CCCZ mirror ($144\times$) \\
\texttt{dc3i8rqnj2vs73didjo0} & \texttt{ibm\_fez}      & Two-copy Bell sampling (8q, 12q; swap expectation \& $\mathcal{W}_2$) \\
\texttt{dc3j25lhdvas73bo0v3g} & \texttt{ibm\_fez}      & 30q GHZ phase scan ($V=0.156$), 3x interleaved reps, full 16q stabilizers \\
\texttt{dc3jsoqnj2vs73digg0g} & \texttt{ibm\_fez}      & 16q 3-CCCZ A/B: baseline vs.\ 23q zero-SWAP ancilla+RCCX ($16.77\%$ / $22.98\%$) \\
\midrule
\texttt{bf50c389-9b33-408e-a3af-e166fd284756} & \texttt{IonQ Forte Ent.} & Magic square Row 0 (cols 0--2 parallel 12q, 160 shots; $96.67\%$) \\
\texttt{47a4f5f8-7f4f-44ac-b7c5-54451f67c3a1} & \texttt{IonQ Forte Ent.} & Magic square Row 1 (cols 0--2 parallel 12q, 160 shots; $99.17\%$) \\
\texttt{21d8261f-e395-42d2-82b8-f2b7c3b86c63} & \texttt{IonQ Forte Ent.} & Magic square Row 2 (cols 0--2 parallel 12q, 160 shots; $96.04\%$) \\
\bottomrule
\end{tabular}%
}
\end{table*}

\end{document}